\documentclass[amsmath,superscriptaddress,showpacs,aps,prb,twocolumn]{revtex4-1}
\usepackage[colorlinks,linkcolor=blue,anchorcolor=blue,citecolor=blue,urlcolor=blue]{hyperref}
\usepackage{amsmath}
\usepackage{amssymb}
\usepackage{graphicx}
\usepackage{color}
\usepackage{bm}

\begin{document}
\title{Ferromagnetism and spin excitations in topological Hubbard models with a flatband}
\author{Xiao-Fei Su}
\affiliation{National Laboratory of Solid State Microstructures and Department of Physics, Nanjing University, Nanjing 210093, China}
\affiliation{School of Physics and Electronic Information, Huaibei Normal University, Huaibei 235000, China}
\author{Zhao-Long Gu}
\affiliation{National Laboratory of Solid State Microstructures and Department of Physics, Nanjing University, Nanjing 210093, China}
\author{Zhao-Yang Dong}
\affiliation{Department of Applied Physics, Nanjing University of Science and Technology, Nanjing 210094, China.}
\affiliation{National Laboratory of Solid State Microstructures and Department of Physics, Nanjing University, Nanjing 210093, China}
\author{Shun-Li Yu}
\affiliation{National Laboratory of Solid State Microstructures and Department of Physics, Nanjing University, Nanjing 210093, China}
\affiliation{Collaborative Innovation Center of Advanced Microstructures, Nanjing University, Nanjing 210093, China}
\author{Jian-Xin Li}
\email[]{jxli@nju.edu.cn}
\affiliation{National Laboratory of Solid State Microstructures and Department of Physics, Nanjing University, Nanjing 210093, China}
\affiliation{Collaborative Innovation Center of Advanced Microstructures, Nanjing University, Nanjing 210093, China}
\date{\today}

\begin{abstract}
\par We study the spin-1 excitation spectra of the flatband ferromagnetic phases in interacting topological insulators. As a paradigm, we consider a quarter filled square lattice Hubbard model whose free part is the $\pi$ flux state with topologically nontrivial and nearly-flat electron bands, which can realize either the Chern or $Z_2$ Hubbard model. By using the numerical exact diagonalization method with a projection onto the nearly-flat band, we obtain the ferromagnetic spin-1 excitation spectra for both the Chern and $Z_2$ Hubbard models, consisting of spin waves and Stoner continuum. The spectra exhibit quite distinct dispersions for both cases, in particular the spin wave is gapless for the Chern Hubbard model, while gapped for the $Z_2$ Hubbard model. Remarkably, in both cases, the nonflatness of the free electron bands introduces dips in the lower boundary of the Stoner continuum. It significantly renormalizes the energies of the spin waves around these dips downward and leads to roton-like spin excitations. We elaborate that it is the softening of the roton-like modes that destabilizes the ferromagnetic phase, and determine the parameter region where the ferromagnetic phase is stable.
\end{abstract}
\maketitle

\section{Introduction}
\par Electronic bands with nonzero topological indices reside on the center of a substantial amount of topological phenomena in condensed matter physics \cite{HK_RMP2010,QZ_RMP2011}. It was proposed in a pioneering work by Haldane \cite{H_PRL1988} that a spinless fermionic model on a honeycomb lattice exhibits integer quantum Hall effect \cite{KDP_PRL1980} without an external magnetic field. This model, serving as the first example of Chern insulator, breaks the time reversal symmetry with a complex next-nearest-neighbor hopping and is characterized by a nonzero Chern number \cite{TKNN_PRL1982}. Later, the concept was generalized to time reversal symmetric systems with spin-orbital coupling (SOC), such as the graphene \cite{KM_PRL2005a,KM_PRL2005b} and HgTe/CdTe quantum wells \cite{BHZ_S2006,Ketc_S2007}. The SOC there generates complex hopping terms similar to that proposed by Haldane but with opposite chiralities for electrons with up spins and down spins, resulting in the quantum spin Hall insulator characterized by a $Z_2$ index.

\par The lattice models with nontrivial band topology share much similarity with the two-dimensional electron gas (2DEG) under a strong magnetic field with Landau levels, e.g. the existence of topologically protected gapless edge states \cite{H_PRL1988,L_PRB1981,H_PRB1982,KM_PRL2005a,KM_PRL2005b,BHZ_S2006,Ketc_S2007,YXL_PRL2011,LYGL_PRB2016,LGLW_NJP2017}. Thus more novel phases other than the Chern insulator or $Z_2$ insulator are expected when Coulomb interactions are taken into account, as is similar to the fractional quantum Hall effect \cite{TSG_PRL1982,L_PRL1983} in the 2DEG with Landau levels. However, different from Landau levels, energy bands in lattice models usually have noneligible dispersions, which weakens the effect of Coulomb interactions. Therefore, in recent years, much effort has been devoted to the design and search of tight-binding models that host nearly-flat electron bands with nontrivial topology \cite{TMW_PRL2011,WR_PRB2011,SGKD_PRL2011,WGGS_PRL2011,NSCM_PRL2011,TB_PRB2012,YGSD_PRB2012}. Analogous exotic phases, such as the fractional Chern insulator and fractional topological insulator were numerically verified to emerge in such nearly-flat topological bands when strong Coulomb interactions are turned on \cite{NSCM_PRL2011,SGSS_NC2011,RB_PRX2011,NSRCM_PRB2011,LBFL_PRL2012}.

\par Another involved intriguing phenomenon arising from Coulomb repulsions in flat or nearly-flat bands is the itinerant ferromagnetism \cite{T_PTP1998,T_PRL1992,M_PLA1993,MT_CMP1993,KMTT_EPL2010}. It was proved by Tasaki and Mielke that the ground state of a flat electron band with a filling factor not more than but sufficiently close to $1/2$ is ferromagnetically ordered as long as an infinitesimal onsite Hubbard repulsion is present \cite{T_PRL1992,M_PLA1993,MT_CMP1993}. Afterwards, this ferromagnetism was shown to be stable against small nonflatness of the electron bands if and only if the Hubbard interaction exceeds a critical value \cite{T_PRL1994}. Spin wave excitations over this ferromagnetic ground state were also studied \cite{KA_PRL1994,SGDL_PRB2018} and itinerant topological magnons have been reported quite recently \cite{SGDL_PRB2018}.

\par The interplay between flatband ferromagnetism and nontrivial band topology enriches the related physics. In fact, ferromagnetism is essential in the generation of stable fractional Chern insulators in the proposals where the spin degrees of freedom of electrons are considered \cite{TMW_PRL2011,NSRCM_PRB2011,LBFL_PRL2012}. Furthermore, ferromagnetism can also lead to possible high-temperature quantum anomalous Hall effect (QAHE) when the nearly-flat topological band is half-filled \cite{NSRCM_PRL2012}. In this paper, we study the ferromagnetism and  spin excitations from the ferromagnetic ground state in nearly-flat topological bands. As a paradigm, we consider a square lattice Hubbard model whose free part is a $\pi$ flux model with topologically nontrivial and nearly-flat electron bands. Depending on the nearest-neighbor hopping, the model Hamiltonian either explicitly breaks the time-reversal symmetry but preserves the spin $SU(2)$ rotation symmetry (Chern Hubbard model), or preserves the time-reversal symmetry but explicitly breaks the spin $SU(2)$ rotation symmetry ($Z_2$ Hubbard model). When the model is quarter filled (or correspondingly, the lower nearly-flat band is half filled), the ground state is spin fully polarized due to the ferromagnetism and exhibits QAHE because of the nonzero Chern number of a single-spin band. Consequently, the charge excitations are gapped and the low energy physics is dominated by the one-spin-flip excitations. These spin-1 excitations have been studied by a generalized bosonization scheme where the interacting fermionic model is mapped to a free bosonic model describing spin-wave excitations at the harmonic approximation \cite{DG_PRB2015}. The ferromagnetism was shown to be stable against such spin wave excitations, which are gapless in the Chern Hubbard model and gapped in the $Z_2$ Hubbard model. However in this bosonization scheme, the free part of the electron model plays no role in the spin wave excitations other than contributing a global constant, suggesting that it should fail due to the competition between the kinetic energy and potential energy of electrons when the nonflatness of the electron bands is not negligible \cite{T_PRL1994}. In fact, in a strictly local periodic tight-binding model, an energy band with a nonzero Chern number cannot be exactly flat \cite{CMST_JPMT2014}. Therefore it remains an open question on whether the ground state is stable against the spin-1 excitations and how the nonflatness of the electron bands manifests itself in the spin-1 excitation spectra in such models.

\par To elucidate these questions, we adopt the numerical exact diagonalization method with a projection onto the lower nearly-flat band to take close investigations on the spin-1 excitations of the models. A critical magnitude of the Hubbard interaction is found for both the Chern Hubbard model and $Z_2$ Hubbard model, below which the ferromagnetic phase is unstable. Furthermore, the spin-1 excitation spectra are shown to consist of collective modes (spin waves) and individual modes (Stoner continuum). For the Chern Hubbard model, the spin wave is gapless while for the $Z_2$ Hubbard model, the spin wave is gapped. Remarkably, for both cases, the nonflatness of the free electron bands introduces dips of the lower boundary of the Stoner continuum, and significantly renormalizes the energies of the collective modes around these dips downward, which leads to roton-like spin wave excitations. With the increase of this nonflatness, the energy of the induced roton-like modes goes down and finally touches zero, which results in the destabilization of the ferromagnetic phase. Therefore, we elaborate the mechanism of the instability of this flat-band ferromagnetism as the softening of the emergent roton-like modes with the increase of noflatness. We also make a comparison of our results in the flatband limit with those obtained by the bosonization scheme \cite{DG_PRB2015}. Qualitative agreements are observed for the Chern Hubbard model, yet notable discrepancy appears for the $Z_2$ Hubbard model. We attribute the discrepancy to the multimagnon processes, which are ignored in the bosonization scheme but treated exactly in our method after the projection.

\par The rest of the paper is organized as follows. In Sec. \ref{mm}, we introduce the Chern Hubbard model and $Z_2$ Hubbard model studied in this paper, discuss the nontrivial band topology of their free parts, interpret the emergence of QAHE resulting from the interplay between flatband ferromagnetism and nontrivial band topology, and formulate the exact diagonalization method with a projection onto the lower nearly-flat band on details. In Sec. \ref{nr}, we discuss the phase diagram and elaborate the spin-1 excitation spectra of both models. Section \ref{sd} provides a summary and discussion.

\section{Model and method}\label{mm}
\subsection{Introduction to model}\label{intro_to_model}

\begin{figure}
\includegraphics[width=\columnwidth]{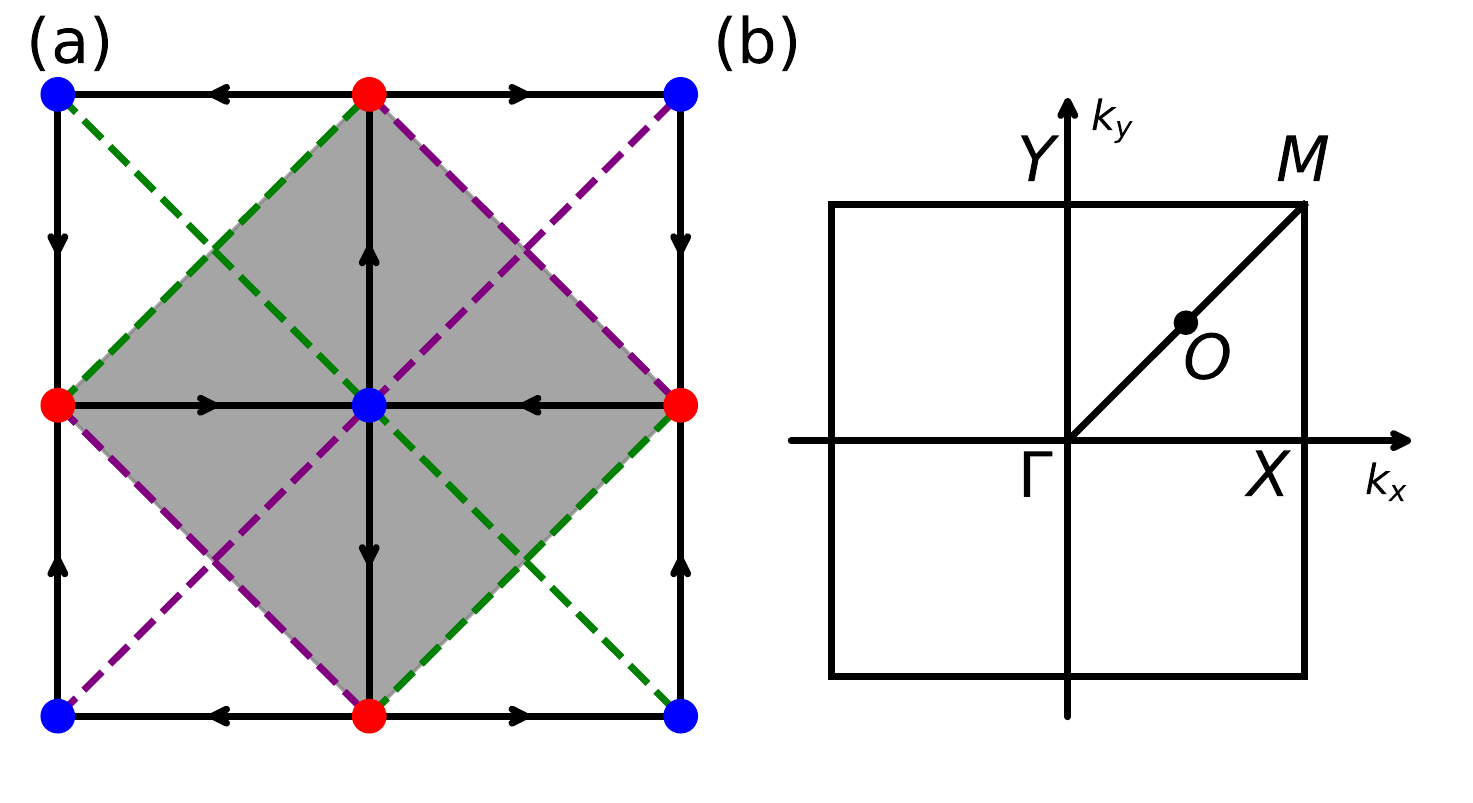}
\caption{(Color online) (a) Schematic representation of the $\pi$-flux model. Blue and red solid circles denote the $A$ and $B$ sublattices, respectively. The nearest-neighbor hopping amplitudes (solid black lines) are equal to $t_{1}\exp(i\alpha_{\sigma}\pi/4)$ (see text) along the direction of the arrows, the next-nearest-neighbor hopping amplitudes are equal to $t_{2}$ (dashed green lines) and $-t_{2}$ (dashed purple lines). The shaded area denotes the unit cell. (b) The first Brillouin zone. The next-nearest-neighbor distance is set to one, so $\Gamma=(0,0)$, $X=(\pi,0)$, $Y=(0,\pi)$, $M=(\pi,\pi)$, $O=(\frac{\pi}{2},\frac{\pi}{2})$.}
\label{lattice}
\end{figure}

\par We consider a generalized $\pi$-flux Hubbard model on the square lattice, whose Hamiltonian can be written as $H=H_0+H_U$, where $H_0$ is the spinfull genralization of the original spinless model proposed in Ref. \onlinecite{NSCM_PRL2011},
\begin{equation}
H_0=\sum_{\langle ij\rangle,\sigma}(t_1^{ij,\sigma}c_{i\sigma}^\dagger c_{j\sigma}+\text{H.c.})
    +\sum_{\langle\langle ij\rangle\rangle,\sigma}(t_2^{ij}c_{i\sigma}^\dagger c_{j\sigma}+\text{H.c.}),
\end{equation}
and $H_U$ is the Hubbard interaction
\begin{equation}
H_U=U\sum_{i}n_{i\uparrow}n_{i\downarrow}.
\end{equation}
Here, $c_{i\sigma}^{\dagger}(c_{i\sigma})$ creates (annihilates) a spin $\sigma$ electron at site $i$, $n_{i\sigma}=c^{\dag}_{i\sigma}c_{i\sigma}$ is the particle-number operator, $\langle ij\rangle$ denotes the nearest-neighbor (NN) bonds and $\langle\langle ij\rangle\rangle$ the next-nearest-neighbor (NNN) bonds. As shown in Fig. \ref{lattice}(a), the spin-dependent NN hopping amplitude $t_1^{ij,\sigma}$ and the spin-independent NNN hopping amplitude $t_2^{ij}$ are given by
\begin{equation}
t_1^{ij,\sigma}=t_1\exp\left(i\delta_1^{ij}\alpha_\sigma\pi/4\right)
\end{equation}
and
\begin{equation}
t_2^{ij}=t_2\delta_2^{ij},
\end{equation}
respectively. Here, $\delta_1^{ij}=+1$ if the NN electron hopping is along the direction of the solid black arrow and $\delta_1^{ij}=-1$ if along the reversed direction. $\delta_2^{ij}=+1$ if the NNN electron hopping is along the dashed green lines and $\delta_2^{ij}=-1$ if along the dashed purple lines. The spin-dependent phase factor $\alpha_\sigma$ breaks the time-reversal symmetry but preserves the spin $SU(2)$ rotation symmetry if $\alpha_\uparrow=\alpha_\downarrow=+1$, whereas it preserves the time-reversal symmetry but breaks the spin $SU(2)$ rotation symmetry if $\alpha_\uparrow=+1$ and $\alpha_\downarrow=-1$.

\par Due to the complex NN hopping, each electron will acquire a $\pi$ phase as it hops around a plaquette along the direction of the black arrows as indicated in Fig. \ref{lattice}(a). Therefore, $H_0$ describes free electrons hopping on a square lattice in the presence of a fictitious staggered $\pi$-flux pattern \cite{WWZ_PRB1989}. For the time-reversal-symmetry-breaking case, $\alpha_\uparrow=\alpha_\downarrow$, the fluxes experienced by spin-up electrons and spin-down electrons are the same, while for the time-reversal-symmetry-preserving case, $\alpha_\uparrow=-\alpha_\downarrow$, they are opposite.

\subsection{Topology of free term $H_0$}\label{topo_free_part}
\par Gapped noninteracting fermionic systems can be topologically classified by their Hamiltonians in the momentum space in the presence of symmetries \cite{SRFL_PRB2008}. After the Fourier transformation, the free part $H_0$ of our model reads
\begin{equation}
H_0=\sum_{\mathbf{k}\sigma}\psi^{\dagger}_{\mathbf{k}\sigma}h_{\mathbf{k}\sigma}\psi_{\mathbf{k}\sigma},
\end{equation}
where $\psi^{\dagger}_{\mathbf{k}\sigma}=(c^{\dagger}_{A\mathbf{k}\sigma},c^{\dagger}_{B\mathbf{k}\sigma})$ and
\begin{equation}
h_{\mathbf{k}\sigma}={\mathbf{D}_{\mathbf{k}\sigma}}\cdot\bm{\tau}.
\end{equation}
Here $\bm{\tau}=(\tau_{1},\tau_{2},\tau_{3})$ is a $2\times2$ matrix vector, $\tau_{1},\tau_{2},\tau_{3}$ are the three Pauli matrices for the sublattice degrees of freedom. The components of $\mathbf{D}_{\mathbf{k}\sigma}$ are given by
\begin{equation}
\begin{aligned}
&D_{1,\mathbf{k}}=2\sqrt{2}t_{1}\cos\frac{k_{x}}{2}\cos\frac{k_{y}}{2},\\
&D_{2,\mathbf{k}}=2\sqrt{2}t_{1}\alpha_\sigma\sin\frac{k_{x}}{2}\sin\frac{k_{y}}{2},\\
&D_{3,\mathbf{k}}=2t_{2}(\cos k_{x}-\cos k_{y}).
\end{aligned}
\end{equation}
$H_0$ can be diagonalized with the transformation
\begin{equation}
\begin{aligned}
&c_{A\mathbf{k}\sigma}=\mu_{1,\mathbf{k}\sigma}d_{\mathbf{k}\sigma}+\mu_{2,\mathbf{k}\sigma}f_{\mathbf{k}\sigma},\\
&c_{B\mathbf{k}\sigma}=\mu^{\ast}_{1,\mathbf{k}\sigma}f_{\mathbf{k}\sigma}-\mu^{\ast}_{2,\mathbf{k}\sigma}d_{\mathbf{k}\sigma},\\
\end{aligned}
\end{equation}
where
\begin{equation}
\begin{aligned}
&\mu_{1,\mathbf{k}\sigma}=\frac{D_{1,\mathbf{k}}-i\alpha_\sigma D_{2,\mathbf{k}}}{\sqrt{2D_{\mathbf{k}}(D_{\mathbf{k}}+D_{3,\mathbf{k}})}},\\
&\mu_{2,\mathbf{k}\sigma}=\frac{D_{\mathbf{k}}+D_{3,\mathbf{k}}}{\sqrt{2D_{\mathbf{k}}(D_{\mathbf{k}}+D_{3,\mathbf{k}})}},\\
\end{aligned}
\end{equation}
with $D_{\mathbf{k}}=\sqrt{D^2_{1,\mathbf{k}}+D^2_{2,\mathbf{k}}+D^2_{3,\mathbf{k}}}$. The diagonalized $H_0$ is given by
\begin{equation}
H_{0}=\sum_{\mathbf{k}\sigma}\varepsilon_{d}(\mathbf{k})c^{\dagger}_{\mathbf{k}\sigma}c_{\mathbf{k}\sigma}
      +\sum_{\mathbf{k}\sigma}\varepsilon_{f}(\mathbf{k})f^{\dagger}_{\mathbf{k}\sigma}f_{\mathbf{k}\sigma},
\end{equation}
where $\varepsilon_{d}(\mathbf{k})=-D_{\mathbf{k}}$, $\varepsilon_{f}(\mathbf{k})=D_{\mathbf{k}}$. It can be seen that there exists a gap between the $d$ band and $f$ band when $t_1\ne0$ and $t_2\ne0$.
\par When there is a gap between the $d$ band and $f$ band, these bands can be shown to be topologically non-trivial by calculating their Chern numbers \cite{TKNN_PRL1982} (for the time-reversal-symmetry-breaking case) or $Z_2$ indices \cite{KM_PRL2005b,SWSH_PRL2006} (for the time-reversal-symmetry-preserving case). The Chern number for a single spin component of the $d$ band or the $f$ band can be expressed in terms of the coefficients $D_{i,\mathbf{k}}$ \cite{HK_RMP2010,QZ_RMP2011},
\begin{equation}
C_{\sigma}^{d/f}=\pm\frac{1}{4\pi}\int_{BZ}d^{2}k\;
    \hat{\mathbf{D}}_{\mathbf{k}\sigma}\cdot(\partial_{k_{x}}\hat{\mathbf{D}}_{\mathbf{k}\sigma}\times\partial_{k_{y}}\hat{\mathbf{D}}_{\mathbf{k}\sigma})
    =\pm\alpha_\sigma,
\end{equation}
with $\hat{\mathbf{D}}_{\mathbf{k}\sigma}\equiv\mathbf{D}_{\mathbf{k}\sigma}/D_{\mathbf{k}}$.
\par When the system breaks the time-reversal symmetry, i.e. $\alpha_\uparrow=\alpha_\downarrow=1$, $C_\uparrow^d=C_\downarrow^d=1$ and $C_\uparrow^f=C_\downarrow^f=-1$, the total Chern number of the $d$ band is $C^d=C_\uparrow^d+C_\downarrow^d=2$. Therefore, the ground state of $H_0$ will be a noninteracting Chern insulator and exhibits quantum anomalous Hall effect(QAHE) when the lower $d$ band is fully filled. When the system preserves the time-reversal symmetry, i.e. $\alpha_\uparrow=-\alpha_\downarrow=1$, $C_\uparrow^d=-C_\downarrow^d=1$ and $C_\uparrow^f=-C_\downarrow^f=-1$, the total Chern number of the $d$ band is zero. However, the $Z_2$ index, which is defined as
\begin{equation}
\nu=\frac{1}{2}(C_{\uparrow}-C{\downarrow})\mod2,
\end{equation}
of the $d$ band is $1$ and nontrivial. As a consequence, the ground state of $H_0$ will be a noninteracting $Z_2$ insulator and exhibits quantum spin Hall effect(QSHE) when the lower $d$ band is fully filled.

\subsection{Emergence of QAHE in half-filled nearly-flat topological bands}\label{emerg_qahe}
\par For a free fermionic system hosting an energy band with a nonzero Chern number or $Z_2$ index, the distinguished phenomenon resulting from this nontrivial band topology, such as QAHE or QSHE, only manifests itself when the topological band is fully filled. At any fractional filling, the ground state of such a system will be a trivial metal. Intriguingly, when the Coulomb interactions between electrons is introduced, the physics of the nontrivial band topology becomes more involved, especially when the band is nearly flat so that the effects of the Coulomb interactions are highly enhanced. Combined with strong Coulomb interactions, nontrivial topological phases can emerge from fractionally filled topological bands. In this article, we are interested in half-filled strongly-correlated nearly-flat topological bands where QAHE can arise\cite{NSCM_PRL2011,DG_PRB2015}. The essence for the occurrence of this nontrivial phase is the emergence of itinerant ferromagnetism on nearly-flat bands\cite{T_PTP1998}, which fully polarizes all electron spins. Therefore, only one spin component of the topological band will be fully filled exactly, which leads to QAHE due to the nonzero Chern number of that spin component of the band.

\par The Chern Hubbard model and $Z_2$ Hubbard model on square lattice described above serve as the paradigm, where half filling of the lower electron band corresponds to quarter filling of the whole system because of the existence of $AB$ sublattices. When $t_2/t_1$ takes values in a selected region, the $d$ band and $f$ band are quite flat in that the flatness ratio $\Delta/W$, which is defined as the ratio of the gap $\Delta$ between these two bands to the bandwidth $W$ of the lower band, can be as large as $4.83$ \cite{NSCM_PRL2011,DG_PRB2015}. In the next subsection, we will introduce the exact diagonalization method with a projection onto the lower nearly-flat electron band to study the spin-1 excitations of the ferromagnetic phases in these two topological Hubbard models.

\subsection{Exact diagonalization with projection}\label{exact_diag_proj}
\par Exact diagonalization method with a projection onto the low-energy Hilbert space has been widely applied to systems that host flat or nearly-flat energy bands \cite{NSCM_PRL2011,RB_PRX2011,NSRCM_PRB2011,SGDL_PRB2018,NSRCM_PRL2012}. This approach applies when the energy gap between the flat or nearly-flat band and other bands is larger than the Coulomb interaction.

\par For the model we study in this article, the relevant low-energy subspace is the lower electron band, i.e. the $d$ band. Let $P$ denote the corresponding projector, then the Hamiltonian after the projection is
\begin{widetext}
\begin{equation}
P^\dagger HP=\sum_{\mathbf{k}\sigma}\varepsilon_d(\mathbf{k})d^{\dagger}_{\mathbf{k}\sigma}d_{\mathbf{k}\sigma}
+\frac{U}{N}\sum_{a=1,2}\sum_{\mathbf{k}\mathbf{k}^{\prime}\mathbf{q}}
(\mu^{\ast}_{a,\mathbf{k}+\mathbf{q}\uparrow}\mu^{\ast}_{a,\mathbf{k}^{\prime}-\mathbf{q}\downarrow}\mu_{a,\mathbf{k}^{\prime}\downarrow}\mu_{a,\mathbf{k}\uparrow})
d^{\dagger}_{\mathbf{k}+\mathbf{q}\uparrow}d^{\dagger}_{\mathbf{k}^{\prime}-\mathbf{q}\downarrow}d_{\mathbf{k}^{\prime}\downarrow}d_{\mathbf{k}\uparrow}.
\end{equation}
\end{widetext}
Let $|\text{FM}\rangle$ denotes the spin-up fully polarized state on the $d$ band,
\begin{equation}
|\text{FM}\rangle=\prod_{\mathbf{k}\in \text{FBZ}}d^{\dagger}_{\mathbf{k}\uparrow}|\text{0}\rangle,
\end{equation}
where FBZ denotes the first Brillouin zone [see Fig. \ref{lattice}(b)] and $|0\rangle$ is the fermion vaccum. Here, $d^{\dag}_{\mathbf{k}\uparrow}$ creates a spin-up electron with momentum $\mathbf{k}$. Then the basis of the spin-1 excitations with a center-of-mass momentum $\mathbf{q}$ over this reference state can be written as
\begin{equation}
|\mathbf{k}_i\rangle_\mathbf{q}=d^{\dagger}_{\mathbf{k}_{i}-\mathbf{q}\downarrow}d_{\mathbf{k}_{i}\uparrow}|\text{FM}\rangle,
\end{equation}
which labels a spin-1 scattering channel with the index $\mathbf{k}_{i}$. The dimension of this Hilbert space scales linearly with respect to the number of electron momentums \cite{SGDL_PRB2018}, so a much larger system can be numerically accessed than the usual exact diagonalization without projection. It enables us to analyze the properties of the spin-1 excitation spectra in detail in the whole first Brillouin zone rather than some restricted discrete points solely. The matrix element of the projected Hamiltonian on this spin-1 excitation basis can be easily obtained after some algebra,
\begin{widetext}
\begin{equation}\label{PHP}
\begin{aligned}
_\mathbf{q}\langle\mathbf{k}_j|P^\dagger HP|\mathbf{k}_i\rangle_\mathbf{q}&=\left[\varepsilon_d(\mathbf{k}_i-\mathbf{q})-\varepsilon_d(\mathbf{k}_i)
+\frac{U}{N}\sum_{a=1,2}\sum_{\mathbf{p}\neq\mathbf{k}_i}\left|\mu_{a,\mathbf{p}\uparrow}\right|^2\left|\mu_{a,\mathbf{k}_{i}-\mathbf{q}\downarrow}\right|^2
\right]\delta_{\mathbf{k}_j,\mathbf{k}_i} \\
& -\frac{U}{N}\sum_{a=1,2}
\mu^{\ast}_{a,\mathbf{k}_{i}\uparrow}\mu_{a,\mathbf{k}_{j}\uparrow}\mu^{\ast}_{a,\mathbf{k}_j-\mathbf{q}\downarrow}\mu_{a,\mathbf{k}_i-\mathbf{q}\downarrow}
\left(1-\delta_{\mathbf{k}_j,\mathbf{k}_i}\right).
\end{aligned}
\end{equation}
\end{widetext}
Here $\delta_{\mathbf{k}_j,\mathbf{k}_i}$ is the Kronecker delta function. Then the full spin-1 excitation spectra can be obtained by the diagonalization of the matrix whose elements are defined by Eq. (\ref{PHP}). It is noted that $|\text{FM}\rangle$ is the true ground state only if the whole spin-1 excitation spectra have no negative energies. Thus we can use this as the criterion to determine the destabilization of the ferromagnetic phase.
\par We also want to give some remarks on the flatband limit in the framework of this method. The flatband-limit Hamiltonian shares the same single-particle eigenfunctions as well as the interaction terms with the original one, yet it has exactly-flat single-particle energy bands\cite{NSCM_PRL2011}. Thus the free part $h^\text{flat}_{\mathbf{k}\sigma}$ of the flatband-limit Hamiltonian can be defined as
\begin{equation}
h^{\text{flat}}_{\mathbf{k}\sigma}=\frac{h_{\mathbf{k}\sigma}}{\left|\varepsilon_{d}(\mathbf{k})\right|}={\hat{\mathbf{D}}_{\mathbf{k}\sigma}}\cdot\mathbf{\tau}.
\end{equation}
Apparently, the eigenfunctions of $h^\text{flat}_{\mathbf{k}\sigma}$ are the same with $h_{\mathbf{k}\sigma}$'s but its energy bands are exactly flat with the eigenvalues being $\pm 1$. To approach this limit, long-range hopping terms in the real space must be included \cite{NSCM_PRL2011}. To reveal the physics related to the nonflatness of the $d$ band, we also calculated the spin-1 excitation spectra in the flatband limit for comparison. This can be done by simply throwing away the $\left[\varepsilon_d(\mathbf{k}_i-\mathbf{q})-\varepsilon_d(\mathbf{k}_i)\right]\delta_{\mathbf{k}_j,\mathbf{k}_i}$ term in Eq. (\ref{PHP}), because of the same single-particle eigenfunctions shared by the flatband-limit Hamiltonian and the original one.

\section{Numerical results}\label{nr}
\begin{figure}
\includegraphics[width=\columnwidth]{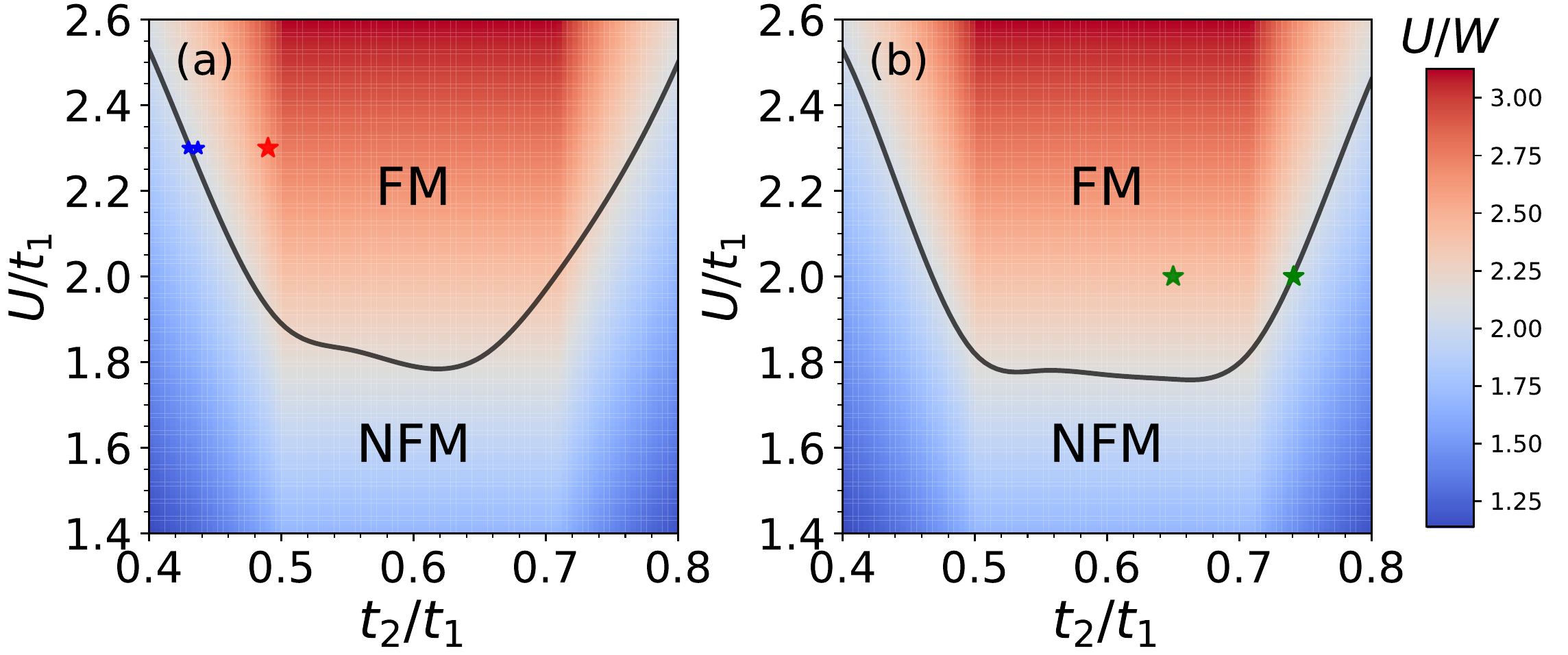}
\caption{(Color online) Phase diagrams of the quarter-filled (a) Chern Hubbard model and (b) $Z_2$ Hubbard model. The colormap represents the ratio $U/W$ of the lower electron band, where $U$ is the Hubbard interaction strength and $W$ is the lower electron bandwidth. Red star in (a) marks the parameter used in Figs. \ref{fmcispectrum} and \ref{bspicture}, blue stars in (a) mark the parameters used in Fig. \ref{pbcispectrum}, green stars in (b) mark the parameters used in Fig. \ref{tispectrum}.}
\label{phase}
\end{figure}

\par Before the detailed discussion on the spin-1 excitation spectra of the quarter-filled Chern Hubbard model and $Z_2$ Hubbard model, we present their phase diagrams first, which are shown in Fig. \ref{phase}. Here, the colormap represents the ratio of the Hubbard interaction strength $U$ to the lower electron bandwidth $W$. FM denotes the ferromagnetic phase and NFM denotes the non-ferromagnetic phase. As discussed in Sec. \ref{exact_diag_proj}, the phase boundary is determined by the onset parameter at which the spin-1 excitation spectrum starts to acquire zero energy at a finite momentum. The phase diagrams of these two models appear quite similar. It is obvious that a critical Hubbard interaction strength is needed to maintain the ferromagnetically ordered ground state when the electron band has finite nonflatness. Furthermore, the phase boundaries always lie near the contour line with $U/W=2.0$. Qualitatively speaking, these observations are expected as a result from the competition between the kinetic energy and potential energy of the electrons, as a fully spin-polarized state minimizes the energy of Hubbard interactions but cost more energy when the electron band disperses. In the following, we will elaborate the spin-1 excitation spectra of these models.

\subsection{Chern Hubbard model}\label{CHM}
\begin{figure}
\includegraphics[width=\columnwidth]{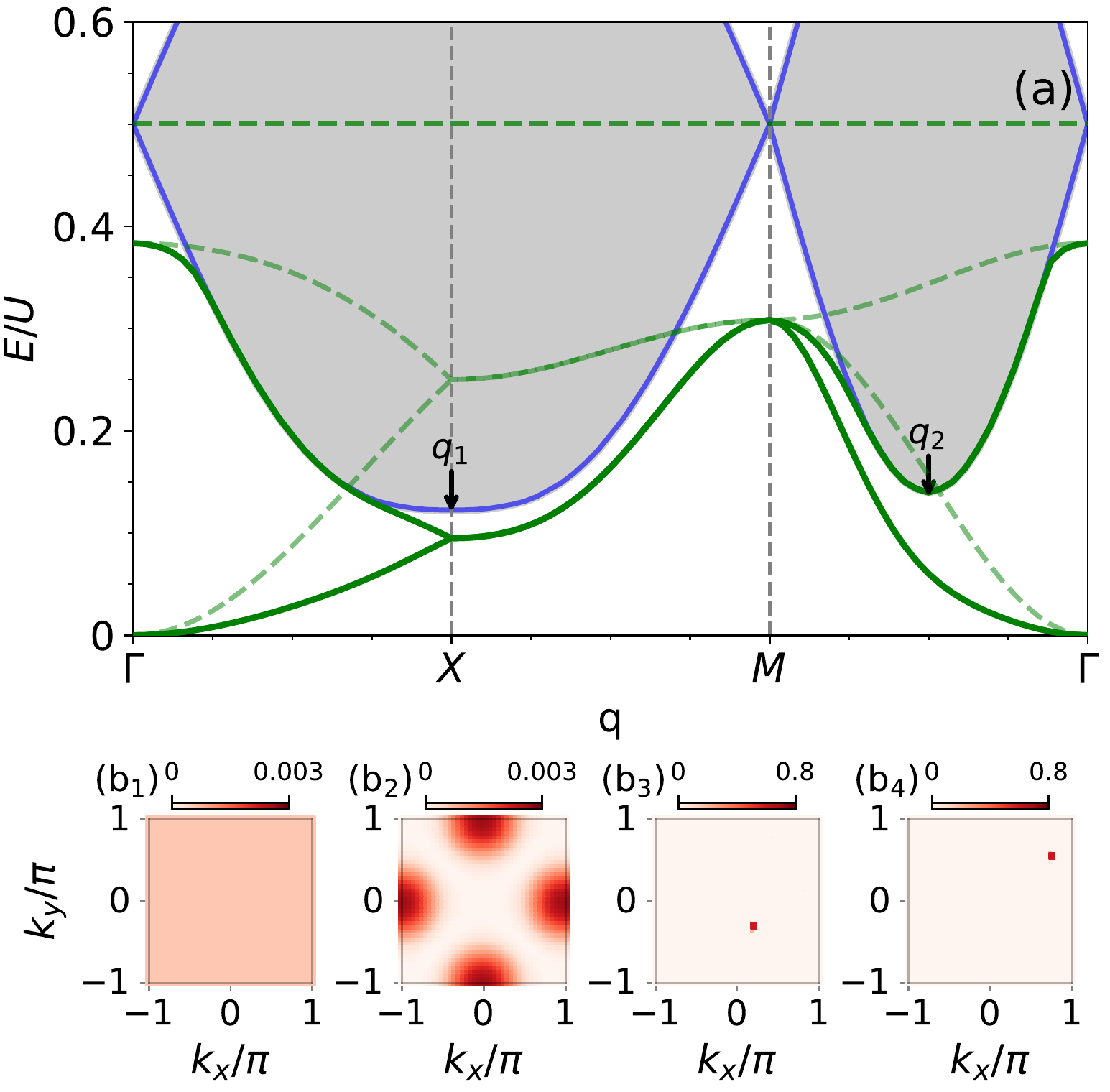}
\caption{(Color online) (a) Spin-1 excitation spectra of the Chern Hubbard model in the FM phase. Green solid lines denote the spin waves and the grey region denotes the Stoner continuum. Blue solid lines denote the upper and lower boundaries of the Stoner continuum determined by the band-splitting picture shown in Fig. \ref{bspicture}(a). Black arrows mark the local minima of the Stoner continuum. (b$_1$-b$_4$) Spectral weights for the lowest four eigen levels of the spin-1 excitation spectra with $\mathbf{q}=(0,0)$. The parameters are $t_1=1.0$, $t_2=0.490$ and $U=2.3$. Green dashed lines in (a) represent the corresponding spectra in the flatband limit.}
\label{fmcispectrum}
\end{figure}

\par In this subsection, we focus on the quarter-filled Chern Hubbard model. In Fig. \ref{fmcispectrum}(a), we present the spin-1 excitation spectra in the FM phase along a high symmetry path in the first Brillouin zone with the parameters marked by the red star in Fig. \ref{phase}(a). It is shown that there are two kinds of obviously different excitations: one contains two low-lying modes labeled by the green solid lines exhibiting well defined band structures,  and the other forms a high-energy continuum as labeled by the shaded area. We find that, for these two kinds of modes, the patterns of the contributions from each spin-1 particle-hole scattering channel of electrons, which are embodied in the eigenvectors $\Psi_{\mathbf{q}}(\mathbf{k}_i)$ of Eq. (\ref{PHP}), are quite different. In Fig. \ref{fmcispectrum}(b), the spectral weights of each scattering channel, i.e. $|\Psi_{\mathbf{q}}(\mathbf{k}_i)|^2$ as a function of $\mathbf{k}_i$, for the four lowest levels with the center-of-mass momentum $\mathbf{q}=(0,0)$ ($\Gamma$ point) are shown. It is clear that for the modes in the green solid lines [Fig. \ref{fmcispectrum}($\text{b}_1$) and Fig. \ref{fmcispectrum}($\text{b}_2$)], the spectral weights come from a quite broad range of scattering channels; indeed, the spectral weight for the lowest mode at $\Gamma$ point is even homogeneous for all scattering channels. However, for those in the shaded area [Fig. \ref{fmcispectrum}($\text{b}_3$) and Fig. \ref{fmcispectrum}($\text{b}_4$)], the spectral weights almost come from a single scattering channel. Therefore, the modes in the green solid lines are collective spin-1 excitations and are identified as the spin waves, while the modes in the shaded area are individual spin-1 excitations and are identified as the Stoner continuum. The spin waves consist of an acoustic branch and an optical branch due to the $AB$ sublattice of the model. The acoustic branch is gapless at the $\Gamma$ point and disperses quadratically away from the $\Gamma$ point, which is the character of a ferromagnetic excitation with a Goldstone mode as the spin-fully polarized ground state spontaneously breaks the spin $SU(2)$ rotation symmetry. Additionally, the acoustic and optical branches are degenerate along the $X$-$M$ path.

\par In the flatband limit, the spin-1 excitation spectra as shown in Fig. \ref{fmcispectrum}(a) as the green dashed lines show several similarities with the dispersive case, including the constitution of the spin waves and Stoner continuum exhibiting as a flat line at $E=0.5U$, the gapless acoustic spin wave and its quadratical dispersion  near the $\Gamma$ point, and the degeneracy between the acoustic and optical branches along the $X$-$M$ path. Comparing results in the flatband limit with those obtained by the generalized bosonization method at the harmonic approximation [see Fig. 4 in Ref. \cite{DG_PRB2015}], one will find that two results are qualitatively quite consistent. In the scheme adopted in Ref. \cite{DG_PRB2015}, the ignorance of the interactions between magnons suggests that only the single-magnon excitations are captured. So, this consistence indicates that the interactions between magnons shows a negligible effect in the flatband limit. However, when the free band becomes dispersive, considerable distinct features appears. Now, the spin-wave energies around the $X=(0,\pi)$ and $O=(\pi/2,\pi/2)$ points are strongly suppressed, as a result two local minima are formed at $X$ and $O$ which can be ascribed to the emergence of the roton-like spin-wave excitations. Furthermore, the continuum exhibiting as a straight line in the flatband limit extends to a large grey region shown in Fig. \ref{fmcispectrum}(a) with dispersive boundaries, and the local minima of its lower boundary coincide with those in the spin-wave dispersion, as indicated by the black arrows in Fig. \ref{fmcispectrum}(a). This observation shows that it is the extending of the Stoner continuum arising from the nonflatness of the $d$ band that pushes down the spin-wave excitations. In particular, it suggests that the couplings between the individual excitations and spin waves would be responsible for the emergence of the roton-like spin waves.

\begin{figure}
\includegraphics[width=\columnwidth]{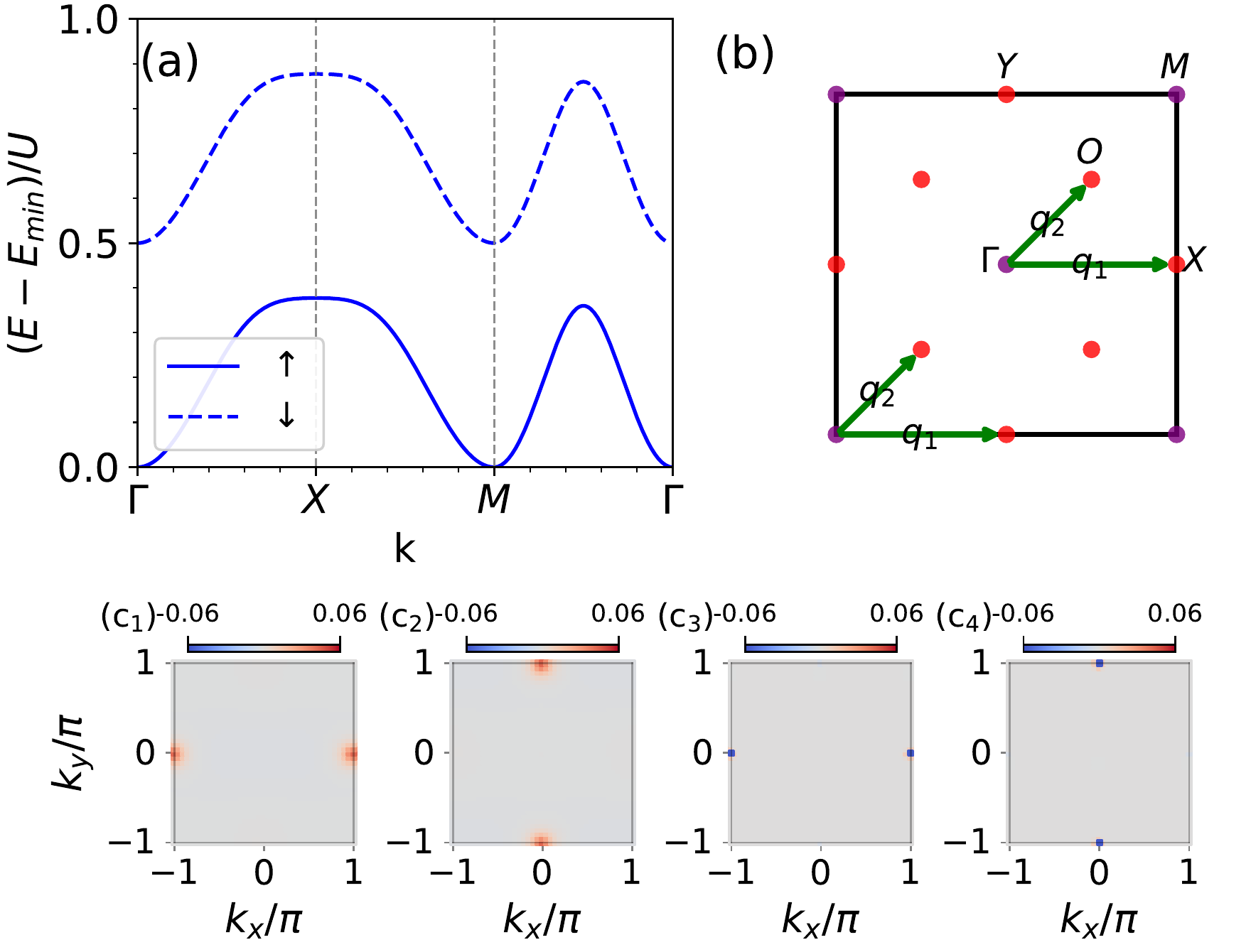}
\caption{(Color online) (a) Illustration of the splitting of the lower electron bands for up spins (blue solid line) and down spins (blue dashed line) in the presence of ferromagnetism of the Chern Hubbard model. (b) Positions of the local maxima (red solid circles) of the spin-up bands and local minima (purple solid circles) of the spin-down bands in the first Brillouin zone. Green arrows denote the corresponding scattering channels of the minima of the Stoner continuum marked in Fig. \ref{fmcispectrum}(a). (c$_1$-c$_4$) Spectral weights subtracted by those in the flatband limit for the lowest four eigen levels of the spin-1 excitation spectra with $\mathbf{q}=(\pi,0)$.}
\label{bspicture}
\end{figure}

\par The formation of the dispersive boundary of the Stoner continuum can be understood by a simple band-splitting picture as shown in Fig. \ref{bspicture}(a). In the presence of ferromagnetism, the spin-down and spin-up electron bands are split in energy which is roughly proportional to $U\langle m \rangle$, with $U$ the Hubbard interaction and $\langle m \rangle$ the average magnetic moment per site. For a quarter-filled electron model with two inequivalent sublattices, $\langle m \rangle=1/2$. Consequently, an individual spin-1 excitation, i.e., a mode in the Stoner continuum, corresponds to the excitation of an electron from the fully-filled spin-up band to the empty spin-down band with an energy proportional to the difference in the initial and final states plus the band splitting $U/2$.
Indeed, the $\delta$ function-like distribution of the spectral weights of a mode in the Stoner continuum [as has be seen in Fig. \ref{fmcispectrum}($\text{b}_3$) and Fig. \ref{fmcispectrum}($\text{b}_4)$] implies that its excitation energy can be approximated by the corresponding diagonal term in Eq. (\ref{PHP}). Numerically, we also find that the term $\frac{U}{N} \sum_{a=1,2} \sum_{\mathbf{p}\neq\mathbf{k}_i} \left| \mu_{a,\mathbf{p}\uparrow} \right|^2 \left| \mu_{a,\mathbf{k}_{i}-\mathbf{q}\downarrow} \right|^2 \sim \frac{U}{2}$. Therefore, the excitation energies of the Stoner continuum are roughly $\varepsilon_{d\downarrow} (\mathbf{k}_i-\mathbf{q}) + \frac{U}{2} -\varepsilon_{d\uparrow} (\mathbf{k}_i)$, which is consistent with the above analysis. The lower and upper boundaries of the Stoner continuum determined in this way are plotted as the blue solid lines in Fig. \ref{fmcispectrum}(a), which fits extremely well with the true boundaries of the shaded area, thus verifies the validity of this simple argument.

\par With this picture, the position of a local boundary bottom of the Stoner continuum is determined by the transferred momentum of the particle-hole scattering between a spin-up maximum and a spin-down minimum. Thus, in Fig. \ref{bspicture}(b), we mark the local maxima of the spin-up band with red solid circles and the local minima of the spin-down band with purple ones. One can see that the transferred momenta coincide completely with the positions
 of the local bottoms in Fig. \ref{fmcispectrum}(a) which are indicated here by green arrows.

\par To understand the renormalization of the spin waves observed above, in Fig. \ref{bspicture}(c), we plot the spectral weights of the four lowest levels subtracted by those in the flatband limit at momentum $q=(\pi,0)$ ($X$ point). The differences concentrate around the scattering channels labeled by $\mathbf{k}_i=(\pm\pi,0)$ and $\mathbf{k}_i=(0,\pm\pi)$. However, the results for the collective modes [Fig. \ref{bspicture}($\text{c}_1$) and Fig. \ref{bspicture}($\text{c}_2$)] are positive while those for the individual modes at the Stoner continuum boundary bottom [Fig. \ref{bspicture}($\text{c}_3$) and Fig. \ref{bspicture}($\text{c}_4$)] are negative. Therefore a noticeable amount of spectral weights transfer from the latter to the former, which indicates a strong coupling between the spin waves and the nearby Stoner continuum. This coupling renormalizes the energies of the spin waves and leads to the occurrence of a roton-like local minimum for the optical branch.

\begin{figure}
\includegraphics[width=\columnwidth]{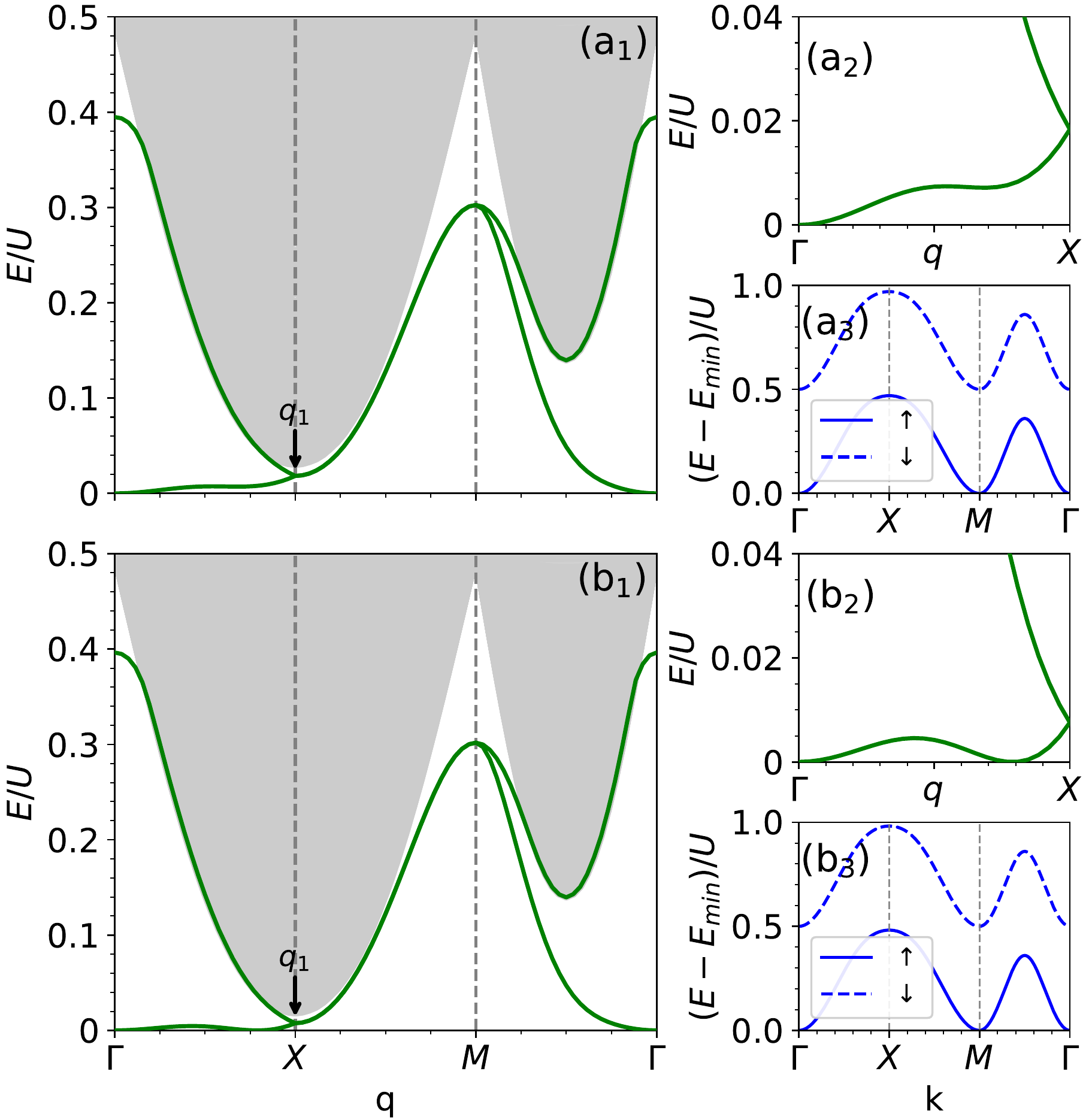}
\caption{(Color online) (a$_1$) and (b$_1$): Spin-1 excitation spectra of the Chern Hubbard model near the FM/NFM phase boundary. Black arrow marks the minimum of the Stoner continuum. (a$_2$) and (b$_2$): Low-energy parts of the spin-1 excitation spectra shown in (a$_1$) and (b$_1$) along the $\Gamma$-$X$ path with an amplified resolution. (a$_3$) and (b$_3$): Illustration of the splitting of the corresponding lower electron bands. $t_2=0.437$ for (a) and $t_2=0.430$ for (b). Other parameters are fixed at $t_1=1.0$, $U=2.3$.}
\label{pbcispectrum}
\end{figure}

\par The emerged roton-like spin waves are essential for the destabilization of the ferromagnetic phase. In Fig. \ref{pbcispectrum}($\text{a}_1$) and \ref{pbcispectrum}($\text{b}_1$), we plot the spin-1 excitation spectra of the model near the FM/NFM phase boundary, with two sets of parameters marked by the blue stars in Fig. \ref{phase}(a). The corresponding illustrations of the band-splitting picture are shown in Fig. \ref{pbcispectrum}($\text{a}_3$) and \ref{pbcispectrum}($\text{b}_3$). Compared with Fig. \ref{fmcispectrum}(a), it can be seen that the increase of the nonflatness leads the boundary bottom marked by $q_1$ to move toward low energies further. In this process, a roton-like local minimum in the acoustic band appears nearby, which can be seen more clearly in Fig. \ref{pbcispectrum}($\text{a}_2$) and \ref{pbcispectrum}($\text{b}_2$). With the increase of the nonflatness of the electron band, the energy of this newly formed roton-like mode goes down and touches zero, which leads to the destabilization of the ground state. From Fig. \ref{pbcispectrum}($\text{a}_3$) and \ref{pbcispectrum}($\text{b}_3$), we can see that the ferromagnetic ground state is quite robust against the spin-1 flips until the top of the spin-up band is approaching near to the bottom of the spin-down band. Considering that the energy splitting of these two bands is $U/2$, the destabilization point would be quite close to $U/2\sim W$.

\subsection{$Z_2$ Hubbard model}\label{ZHM}
\begin{figure}
\includegraphics[width=\columnwidth]{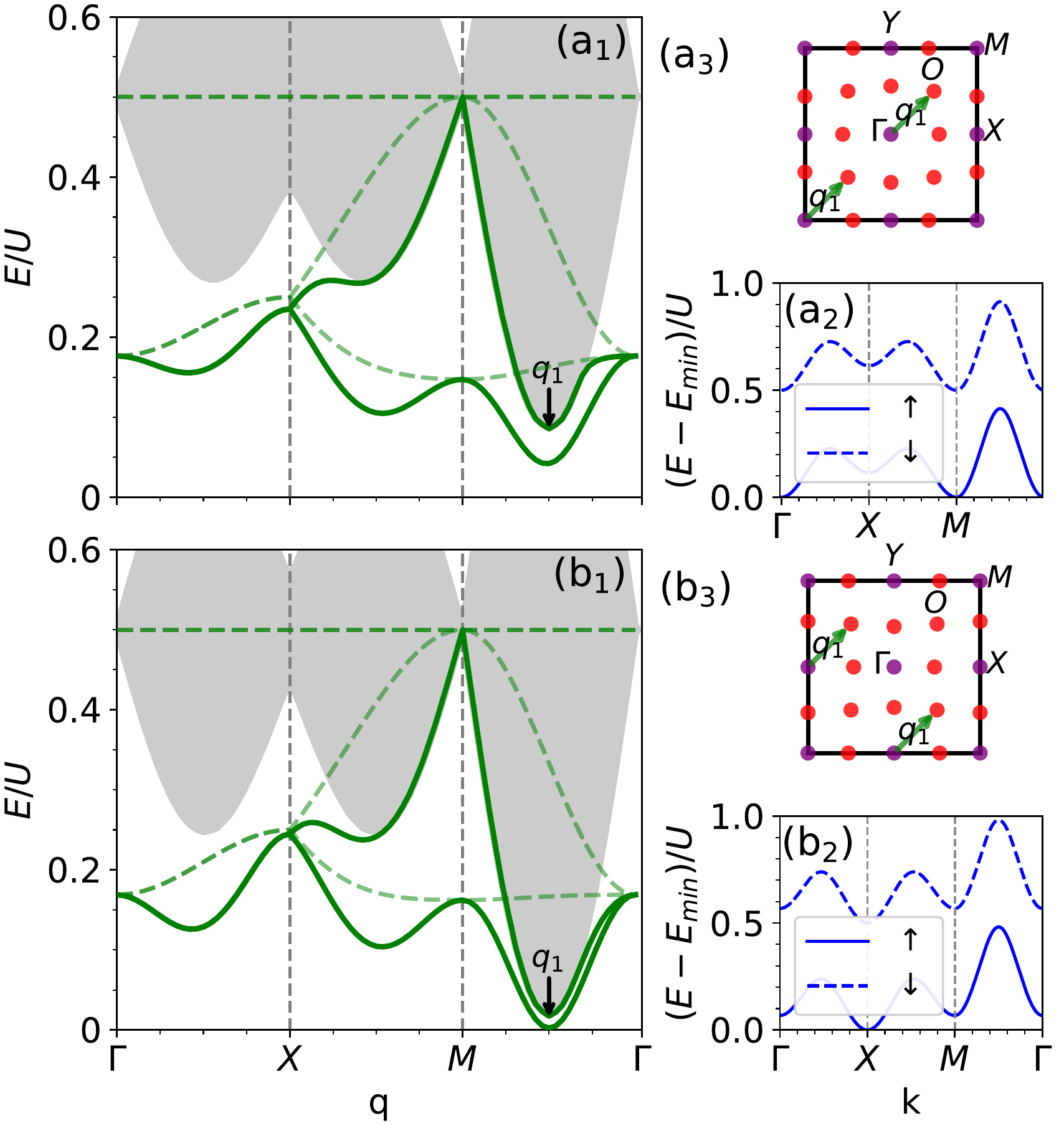}
\caption{(Color online) (a$_1$) and (b$_1$): Spin-1 excitation spectra of the $Z_2$ Hubbard model. Green solid lines denote the spin waves and grey regions denote the Stoner continuum. Green dashed lines represent the corresponding spectra in the flatband limit. Black arrows mark the minima of the Stoner continuum. (a$_2$) and (b$_2$): Illustration of the splitting of the lower electron bands of the $Z_2$ Hubbard model. (a$_3$) and (b$_3$): Positions of the corresponding local maxima (red solid circles) of the spin-up bands and local minima (purple solid circles) of the spin-down bands in the first Brillouin zone. Green arrows denote the corresponding scattering channels of the minima of the Stoner continuum marked in (a$_1$) and (b$_1$). $t_2=0.650$ for (a) and $t_2=0.741$ for (b). Other parameters are fixed at $t_1=1.0$, $U=2.0$.}
\label{tispectrum}
\end{figure}

\par In this subsection, we focus on the quarter-filled $Z_2$ Hubbard model. Its spin-1 excitation spectra along a high symmetry path in the first Brillouin zone with the parameters marked by the green stars in Fig. \ref{phase}(b) are plotted in Fig. \ref{tispectrum}($\text{a}_1$) ($t_2=0.650, U=2.0$) and Fig. \ref{tispectrum}($\text{b}_1$) ($t_2=0.741, U=2.0$), respectively. Same as the case of the Chern insulator discussed in Sec. \ref{CHM}, the low-lying green solid lines are identified as the spin waves (collective modes) and the high-energy shaded areas the Stoner continuum (individual modes). Overall, the whole spectra are different from those in the Chern insulator shown in Fig. \ref{fmcispectrum}(a). In particular, the collective modes here are gapped because the $Z_2$ Hubbard model explicitly breaks the spin $SU(2)$ rotation symmetry and no spontaneous continuous symmetry breaking occurs in the ferromagnetic phase. Besides, the acoustic and optical bands are degenerate along the $\Gamma$-$X$ path, instead of the $X$-$M$ path reported in the subsection \ref{CHM}.

\par The corresponding spectra in the flatband limit are also shown in Fig. \ref{tispectrum}($\text{a}_1$) and Fig. \ref{tispectrum}($\text{b}_1$) as the green dashed lines. Noticeable differences can be found when we compare our result in the flatband limit with that obtained by the generalized bosonization method [see Fig. 5 in Ref. \cite{DG_PRB2015}]. On the one hand, the band bottom of the spin waves in our results lies at the $M$ points while theirs lies at the $\Gamma$ point. On the other hand, the acoustic and optical branches are degenerate along the $\Gamma$-$X$ path, which is same as the results for a dispersive band as discussed in the above. This is in sharp contrast to that obtained in Ref. \cite{DG_PRB2015}, which are degenerate along the $X$-$M$ path. These discrepancies are attributed to the effects of multimagnon processes which are ignored in their scheme. The above observation suggests that the interactions between magnons have an essential effect on the spin excitations in the $Z_2$ Hubbard model, though they have a negligible effect in the Chern Hubbard model as discussed above.

\par Similar to the Chern Hubbard model, the introduction of the dispersion to the flatband leads to the appearance of three local minima (roton-like modes) along the $\Gamma-X-M-\Gamma$ direction for both acoustic and optical bands, respectively. The momentum positions of these minima coincide the corresponding boundary bottoms of the Stoner continuum, suggesting that the renormalization of the spin-wave bands is due to the coupling between the spin waves and Stoner continuum. In the band-splitting picture as proposed above, the position of a local boundary bottom of the Stoner continuum is determined by the transfer momentum of the particle-hole scattering between a spin-up band maximum and a spin-down band minimum. Therefore, we present in Fig. \ref{tispectrum}($\text{a}_2$) and \ref{tispectrum}($\text{b}_2$) the dispersions for the lower electron bands in the $Z_2$ Hubbard model showing the splitting of the spin-up and spin-down bands. The corresponding local maxima (red solid circles) of the spin-up bands and local minima (purple solid circles) of the spin-down bands in the first Brillouin zone are shown in Fig. \ref{tispectrum}($\text{a}_3$) and \ref{tispectrum}($\text{b}_3$). One can see that the momentum positions of the boundary bottoms as marked by the black arrows in Fig. \ref{tispectrum}($\text{a}_1$) and Fig. \ref{tispectrum}($\text{b}_1$) (here only $q_1$ is plotted for illustration) match the transferred momenta between the spin-up and spin-down bands, which gives a strong support of the picture based on the band splitting.

\par From Fig. \ref{tispectrum}($\text{a}_1$) to Fig. \ref{tispectrum}($\text{b}_1$), the ratio of the Hubbard interaction to the lower electron bandwidth $U/W$ decreases. In this process, the roton-like mode with momentum $q_1$ approaches to zero, indicating the destabilization of the ferromagnetic ground state. So, the parameters used to get the result in Fig. \ref{tispectrum}($\text{b}_1$) marks the boundary between a ferromagnetism phase and a nonferromagnetic phase as shown in the phase diagram Fig. \ref{phase}(b). From the band splittings shown in Fig. \ref{tispectrum}($\text{a}_2$) and Fig. \ref{tispectrum}($\text{b}_2$), one can see that the relative energies at the $\Gamma$ and $M$ points increases while that at the $X$ ($Y$) points decreases in this process. Consequently, although the momentum positions of the forementioned boundary bottoms remain the same, the corresponding scattering channels are in fact different, which are shifted from $(0,0)\rightarrow(\frac{\pi}{2},\frac{\pi}{2})$ and $(-\pi,-\pi)\rightarrow(-\frac{\pi}{2},-\frac{\pi}{2})$ to $(-\pi,0)\rightarrow(-\frac{\pi}{2},\frac{\pi}{2})$ and $(0,-\pi)\rightarrow(\frac{\pi}{2},-\frac{\pi}{2})$, as shown in Fig. \ref{tispectrum}($\text{a}_3$) and Fig. \ref{tispectrum}($\text{b}_3$). In fact, we find that the energy of the spin-up band at $O$ almost overlaps with the bottom of the spin-down band at $X$ as shown in Fig. \ref{tispectrum}($\text{b}_2$). This result demonstrates that the ferromagnetic ground state is also quite robust against spin flips until the indirect gap between the spin-down and spin-up bands vanishes, which means the phase boundary of the FM/NFM is also quite close to $U/W=2$ from the discussion in Sec. \ref{CHM}.

\section{Summary and discussion}\label{sd}
\par In summary, we have considered the flatband ferromagnetic phases in a quarter-filled Chern Hubbard model and a quarter-filled $Z_2$ Hubbard model in this paper. By using the numerical exact diagonalization method with a projection onto the lower nearly-flat electron band, we determine the critical Hubbard interaction strength below which the ferromagnetic phase is unstable, and elaborate the ferromagnetic spin-1 excitation spectra of these models. Both spectra consist of collective modes (spin waves) and individual modes (Stoner continuum). For the Chern Hubbard model, the spin wave is gapless while for the $Z_2$ Hubbard model, the spin wave is gapped. Remarkably, in both cases, the nonflatness of the free electron bands introduces dips in the lower boundary of the Stoner continuum. As a result, it renormalizes significantly the energies of the collective modes around these dips through the couplings between spin waves and individual modes in the Stoner continuum, and leads to roton-like spin wave excitations. We find that the destabilization of the ferromagnetic phase arises from the softening of the roton-like mode, whose energy goes down gradually with the increase of this nonflatness.

\par We would like to remark that the downward renormalization of the energies of the collective modes affected by the boundary bottoms of the Stoner continuum shares similarity with that observed in the two-dimensional antiferromagents, where the spin waves are interpreted as bound states of confined spinons and the suppression of energies of spin waves around the continuum bottom is attributed to their couplings to the nearby deconfined spinon continuum\cite{ZFSMC_PRL2006,TS_PRL2013,DMCNTPEMIR_NP2015,SQCCMS_PRX2017,YWDYL_PRB2018}. Intriguing flatband physics arises not only in models with repulsive interactions, but also in those with attractive ones\cite{PT_NC2015,JPVKT_PRL2016,LVPSHT_PRB2017,TLP_arXiv2018}, where the Bardeen-Cooper-Schrieffer superconducting state is the exact ground state in the flatband limit\cite{PT_NC2015,JPVKT_PRL2016}. We hope our methods can be generalized to these systems beyond the flatband limit. In addition, although correlated nearly-flat topological bands are hard to realize in real materials, they are possible to be designed in cold-atom systems. In fact great progress has been made very recently \cite{LCJPS_N2009,AALBPB_PRL2013,MSKBK_PRL2013,JMDLUGE_N2014}. We expect our results will stimulate related investigations in future cold-atom experiments.

\begin{acknowledgments}
\par This work was supported by the National Natural Science Foundation of China (No.11774152, No.11674158) and National Key Projects for Research and Development of China (Grant No. 2016YFA0300401).
\par X.-F. S. and Z.-L. G. contributed equally to this work.
\end{acknowledgments}

\bibliography{ref}
\end{document}